\documentclass{PoS}

\title{Nucleon structure from 2+1f dynamical DWF lattice QCD at nearly physical pion mass}

\ShortTitle{Nucleon structure from 2+1f DWF}

\author{
\speaker{Shigemi Ohta}%\thanks{A footnote may follow.}
\ for RBC and UKQCD Collaborations\\
Institute of Particle and Nuclear Studies, KEK, Tsukuba, Ibaraki 305-0801, Japan\\
Department of Particle and Nuclear Physics, SOKENDAI, Hayama, Kanagawa 240-0193, Japan\\
RIKEN BNL Research Center, Brookhaven National Laboratory, Upton, NY 11973, USA\\
E-mail: \email{shigemi.ohta@kek.jp}
}

\abstract{
Current status of nucleon structure calculations with joint RBC and UKQCD 2+1-flavor dynamical domain-wall fermions (DWF) lattice QCD is reported:
Two ensembles with pion mass of about \(m_\pi=170\) MeV and 250 MeV are used.
The lattice cutoff is set at about 1.4 GeV, allowing a large spatial volume of about \(L=4.6\) fm across while maintaining a sufficiently small residual breaking of chiral symmetry with the dislocation-suppressing-determinant-ratio (DSDR) gauge action.
We calculate all the isovector form factors and some low moments of isovector structure functions.
We confirm the finite-size effect in isovector axialvector-current form factors, in particular the deficit in the axial charge and its scaling in terms of \(m_\pi L\), that we reported from our earlier calculation at heavier pion masses.

\vspace{-171mm}\parbox{\textwidth}{\flushright\large\rm \hfill KEK-TH-1504, RBRC-924}\vspace{171mm}
}

\FullConference{ The XXIX International Symposium on Lattice Field Theory - Lattice 2011\\
July 10-16, 2011\\
Squaw Valley, Lake Tahoe, California}

\begin{document}

\section{RBC/UKQCD (2+1)-flavor dynamical DWF lattice QCD}

During the past several years the RBC and UKQCD collaborations jointly generated three sets of (2+1)-flavor dynamical domain-wall fermions (DWF) lattice QCD ensembles at lattice cut off, \(a^{-1}\), of about 1.7 GeV \cite{Allton:2008pn}, 2.2 GeV \cite{Aoki:2010dy} and 1.4 GeV  \cite{Jung:2010jt,Kelly:LAT2011} respectively:
while the studies using the first two are almost complete and many papers have been published, the study for the last one is ongoing.
In each of these we adjust the strange quark mass to the physical value by reweighting \cite{Jung:2010jt,Aoki:2010dy}.
The degenerate up and down quark masses are set at several values, as light as practical:
In the first two sets it varies in the range corresponding to pion mass from about 420 MeV to 290 MeV.
In the last set there are two ensembles with pion mass of about 250 MeV and 170 MeV.

In contrast to most other lattice fermion methods, the DWF formulation provides good control of chiral and flavor symmetries and non-perturbative renormalization.
This allows us to use meson observables such as pion and kaon mass, \(m_\pi\) and \(m_K\), and decay constants, \(f_\pi\) and \(f_K\), obtained from these ensembles, for a detailed study of meson physics in combined chiral and continuum limit \cite{Aoki:2010dy}: we first use pion, kaon, and \(\Omega\)-baryon mass to set the physical quark mass and lattice scale and then obtain predictions for other observables such as pion and kaon decay constants with a few \% accuracy.
At this level of high accuracy our predictions are now not limited by the statistics but by poor applicability of chiral perturbation or other chiral extrapolation ansatz from the relatively heavy pion mass we used.

Thus the RBC and UKQCD collaborations are now jointly generating a third set of (2+1)-flavor dynamical domain-wall fermions (DWF) lattice QCD ensembles at lower pion mass of 170 MeV and 250 MeV \cite{Jung:2010jt,Kelly:LAT2011}.
This is made possible by using a new combination for gauge action of Iwasaki and dislocation-suppressing-determinant ratio (DSDR) \cite{Vranas:1999rz,Vranas:2006zk,
Renfrew:2009wu} that allows a lower cut off of about 1.4 GeV while keeping the residual breaking of chiral symmetry sufficiently small.
The lower lattice cut off also allows large spatial extent of about 4.6 fm necessary for studying larger hadrons such as nucleon \cite{Yamazaki:2008py,Yamazaki:2009zq,Aoki:2010xg}.
Here I report calculations relevant for nucleon structure using these ensembles.

Such a study, however, requires good understanding of chiral and finite-size corrections to the observables which are unfortunately not well developed for baryons in general.
For this reason, at this stage we are not yet attempting for baryons a combined chiral and continuum limit study that has been done for mesons.
Rather, we are refining our fixed-cutoff study of nucleon to improve our understanding of its chiral behavior.

We published our findings in nucleon structure from the 1.7 GeV ensembles in three publications, Refs.\ \cite{Yamazaki:2008py,Yamazaki:2009zq,Aoki:2010xg}, which we summarize in the following.
As we discovered unacceptably large and distorting dependence on the scaling parameter \(m_\pi L\) of axialvector current form factors, especially at \(m_\pi L = 4.5\), we did not attempt a similar study for our 2.2 GeV ensembles to save our computing resources, and skipped to the new I+DSDR DWF ensembles with lighter pion mass and larger lattice spatial extent, \(L\): the \(m_\pi L\) of these ensembles are respectively at 4.2 and 5.8 and should allow us to better understand the observed dependence.

\begin{figure}[tb]
\begin{center}
\includegraphics[width=0.75\linewidth,clip]{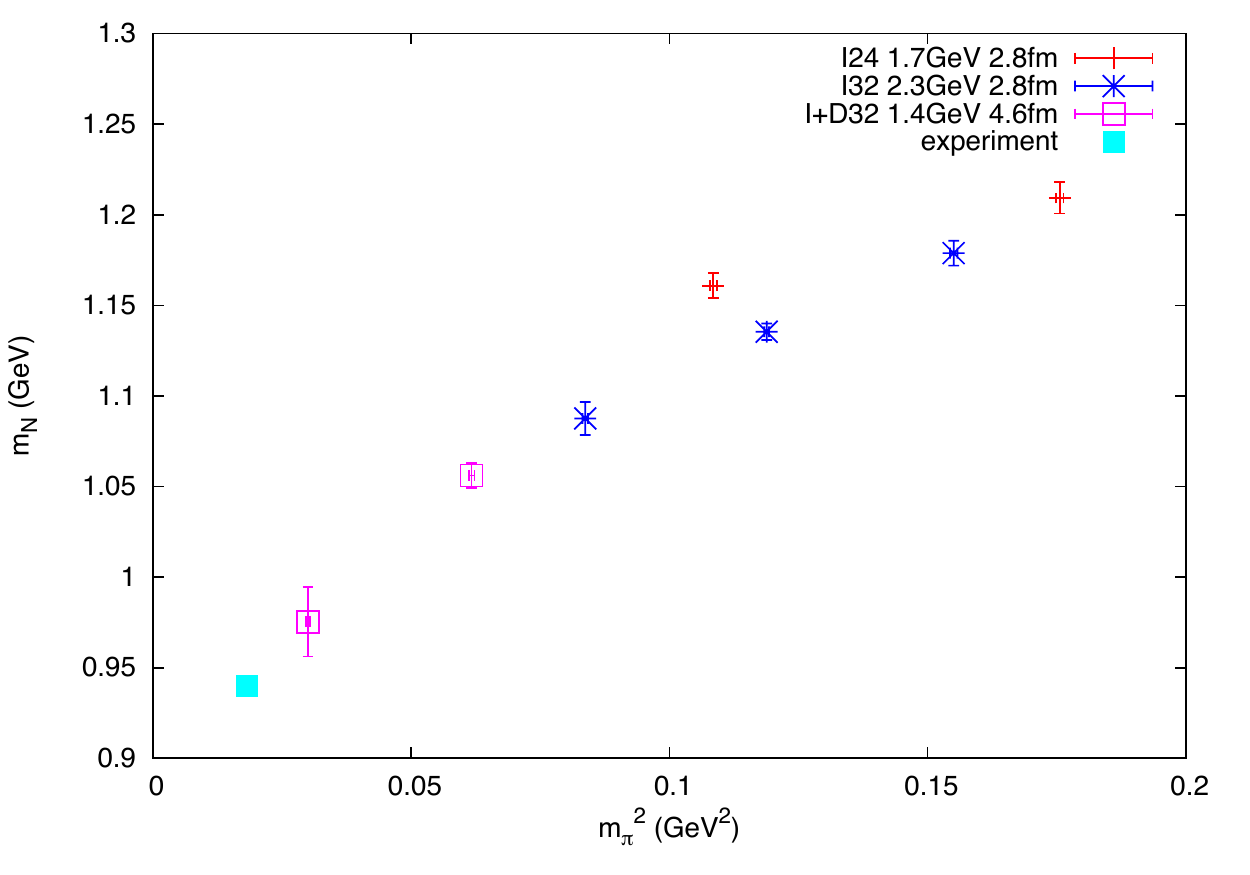}
\end{center}
\caption{
Nucleon mass from the RBC and UKQCD (2+1)-flavor dynamical DWF ensembles plotted against corresponding pion mass squared.  I24 and I32 are from ensembles with simple Iwasaki gauge action while ID32 are with new Iwasaki+DSDR gauge action.
}
\label{fig:masssummary}
\end{figure}

\section{Status at 1.7 GeV: pion cloud surrounding nucleon?}

In a numerical lattice-QCD study \cite{Yamazaki:2008py} the RIKEN-BNL-Columbia (RBC) and UKQCD collaborations reported a strong dependence on pion mass, \(m_\pi\), and lattice spatial extent, \(L\), in the isovector axial charge, \(g_A\), of nucleon.
The dependence is much stronger than had been observed in non-unitary calculations with DWF quarks using either quenched \cite{Sasaki:2003jh} or rooted-staggered fermion \cite{Edwards:2005ym} ensembles.
As the dependence appears to scale with a dimensionless quantity, \(m_\pi L\), the product of calculated pion mass, \(m_\pi\), and the lattice spatial extent, \(L\), a likely explanation for the dependence is that significant part of the nucleon isovector axialvector current is carried by the expected, but never seen, ``pion cloud'' surrounding the nucleon.
If confirmed, this calculation may be the first concrete evidence of such a pion cloud.

Indeed similar strong dependence on pion mass and lattice spatial extent is seen in other axialvector-current form factors \cite{Yamazaki:2009zq} but not in the conserved vector-current ones  \cite{Yamazaki:2009zq}, supporting the pion-cloud interpretation.
However, since the calculations so far have only been carried out at single lattice cut off, \(a^{-1}\), of about 1.7 GeV, two lattice spatial extent, \(L\), of about 1.8 and 2.7 fm, and relatively heavy pion mass down to 330 MeV, it is premature to conclude that the dependence is caused by the pion cloud and so we have discovered it.
This provides a good motivation for the study reported here with pion mass as light as 170 MeV and lattice spatial extent of about 4.6 fm.

For low moments of the isovector structure functions \cite{Aoki:2010xg}, we found the ratio of the quark momentum fraction to the helicity fraction, \(\langle x \rangle_{u-d}/\langle x \rangle_{\Delta u - \Delta d}\), a naturally renormalized quantity on the lattice, in agreement with the experiment, without showing any sign of finite-size effect.
This is in contrast to the case of another naturally renormalized ratio of isovector axial to vector charges, \(g_A/g_V\), which shows strong finite-size dependence as summarized in the above, and consequently underestimates the experiment significantly:
this suggests the low moments of structure functions are not so much affected by the finite lattice volume.
Individual fractions, \(\langle x \rangle_{u-d}\) and \(\langle x \rangle_{\Delta u - \Delta d}\), that had over estimated the experiments at heavier quark mass, showed interesting trending toward experiment at our lightest mass, again without showing any sign of finite-size effect: this also motivates us to study these quantities at lower mass.
To see if this trend is for real is another interesting motivation for the present calculation.

\section{Status at 1.4 GeV}

\begin{figure}[tb]
\begin{center}
\includegraphics[width=0.75\linewidth,clip]{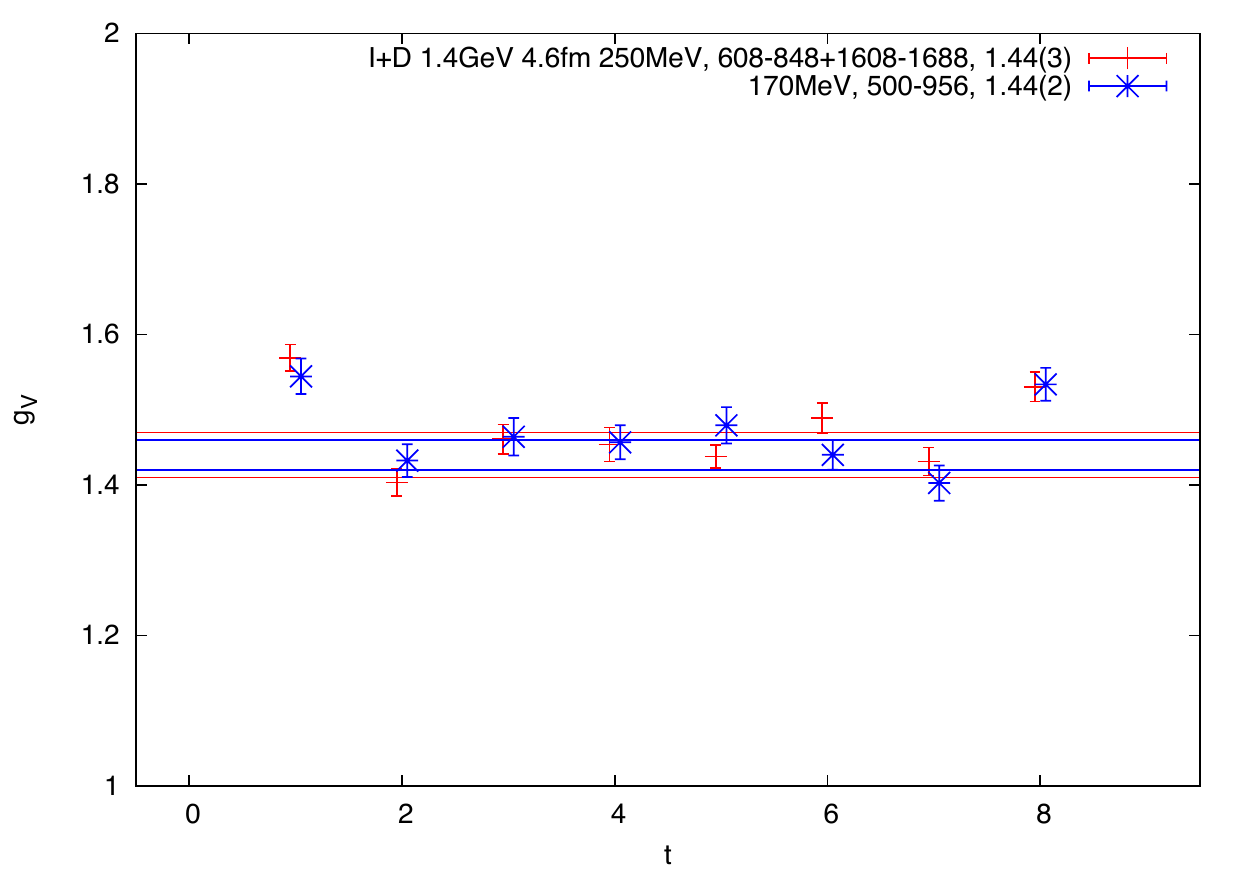}
\end{center}
\caption{
Signals for local-current isovector vector charge: \(g_V = 1.44(3)\) and 1.44(2), is obtained, respectively for the heavy and light ensembles.
These correspond to a non-perturbative renormalization of \(Z_V = g_V^{-1} = 0.69(2)\), in agreement with the axial-current renormalization obtained in the meson sector, up to \(O(a^2)\) correction.
}
\label{fig:gV}
\end{figure}
Some details of these ensembles are reported by C.~Kelly in these proceedings, and we plan to publish a full report soon \cite{Kelly:LAT2011}.
At the time of this symposium, we had about 1900 hybrid Monte Carlo time units for the heavier, \(m_\pi=250\) MeV ensemble and 1200 for the lighter, \(m_\pi=170\) MeV, of which first 600 and 500 respectively are discarded for thermalization.
We analyze configurations separated from each other by eight hybrid Monte Carlo time units. 
For each configuration we calculate four nucleon propagation with sources separated evenly in Euclidean time.
For this report we had analyzed about forty configurations from the heavy ensemble and about sixty from the light respectively.

We use Gaussian smearing \cite{Alexandrou:1992ti, Berruto:2005hg} for nucleon source to optimize the overlap with the ground state and compared the cases for widths 4 and 6 lattice units.
We found width 6 is better for both pion mass \cite{Ohta:2010sr}.
We quote preliminary nucleon mass estimates of 0.721(13) and 0.763(10) lattice units which correspond to about 0.98 and 1.05 GeV with another preliminary estimate for the lattice cut off of 1.368(7) GeV.
As are summarized in Fig.\ \ref{fig:masssummary}, these ensembles are filling the gap toward the physical point.

With the available statistics for separation of eight trajectories, we chose the nucleon source-sink separation of 9 lattice units, or about 1.3 fm in physical units for the present work.
For the expectation values of observables reported here, we use the plateaux between \(t=2\) and 7.
Whether these choices are adequate or not needs further study, which is planned.

Signals for the isovector vector charge, \(g_V\), are shown in Fig.\ \ref{fig:gV}.
As the vector current is conserved, the charge is not affected by excited states nor finite volume, and the signals are very clear: we obtain values of \(g_V = 1.44(3)\) and 1.44(2), respectively for the heavy and light ensembles.
These lead to a non-perturbative renormalization of \(Z_V = g_V^{-1} = 0.69(2)\), in agreement with the axialvector-current renormalization obtained in the meson sector, \(Z_A= 0.6871(4)\) \cite{Kelly:LAT2011}, as is expected up to \(O(a^2)\) correction: the chiral symmetry is well preserved even for the present low lattice cut off of about 1.4 GeV.

\begin{figure}[tb]
\begin{center}
\includegraphics[width=0.75\linewidth,clip]{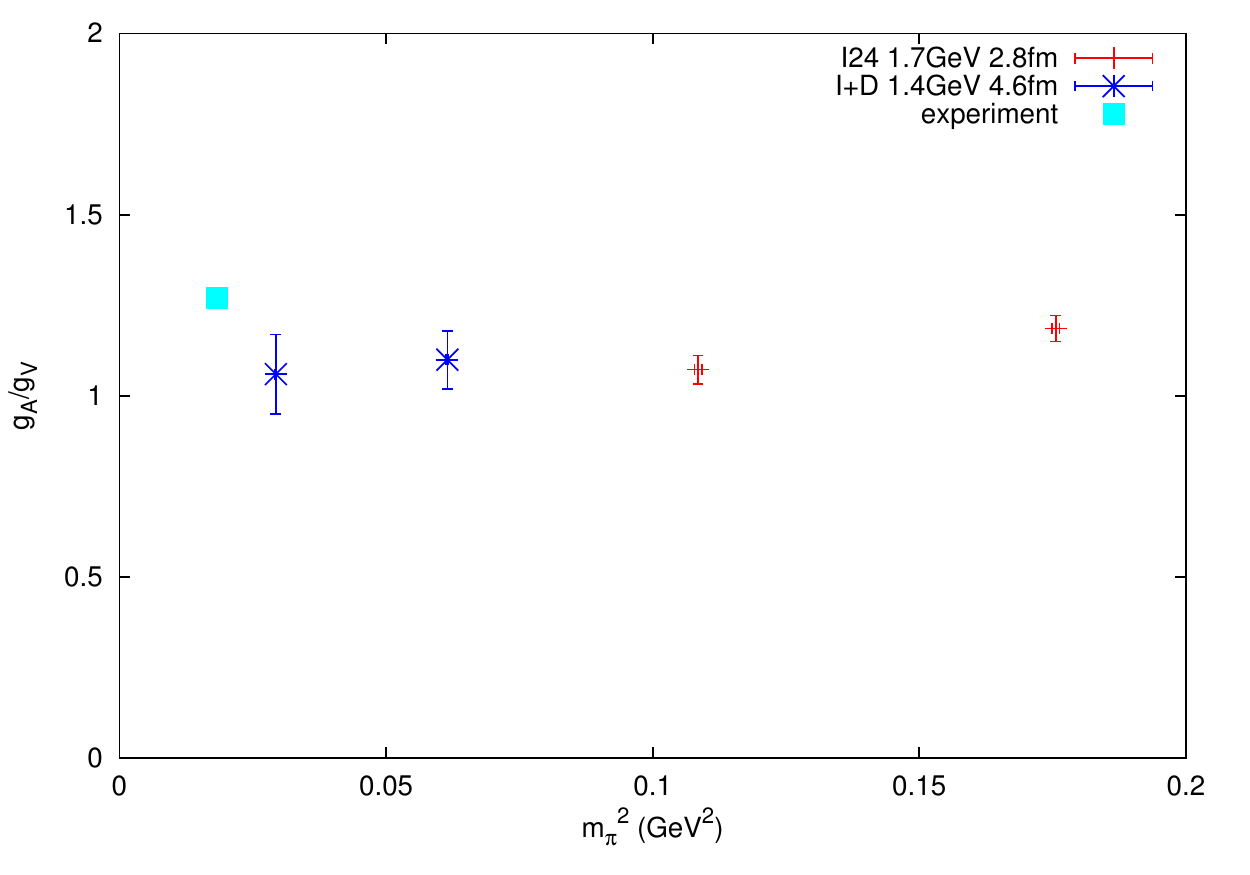}
\end{center}
\caption{
Dependence on pion mass squared, \(m_\pi^2\), of isovector axial to vector charges, \(g_A/g_V\).
The calculated values stay away from the experiment as we set the pion mass lighter.
}
\label{fig:gAgVmpi2}
\end{figure}

Signals for the axial charge itself is much noisier, but the naturally renormalized ratio of the isovector axial to vector charges, \(g_A/g_V\), yields decent signals, 
\begin{figure}[tb]
\begin{center}
\includegraphics[width=0.75\linewidth,clip]{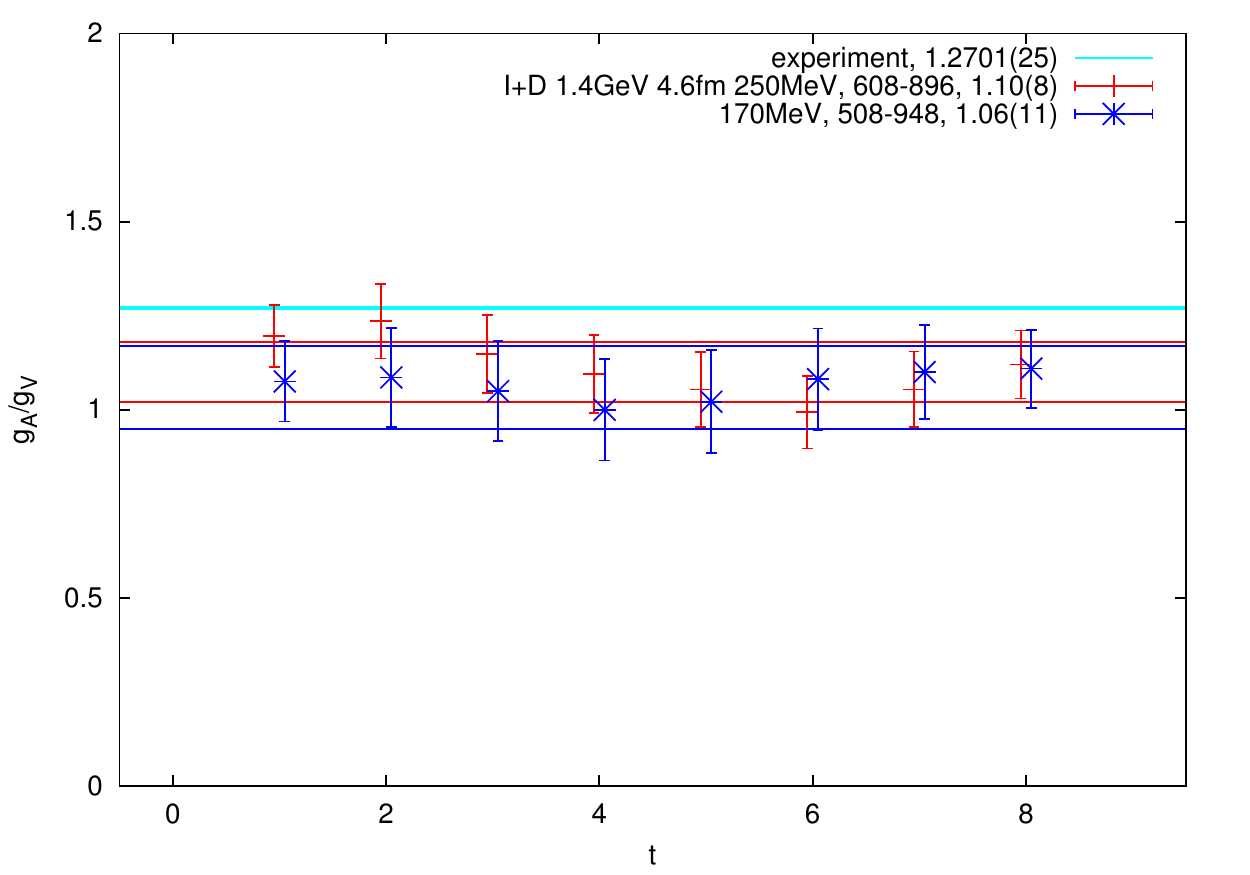}
\end{center}
\caption{
Signals for isovector axial to vector charges, \(g_A/g_V\).}
\label{fig:gAgV}
\end{figure}
as are summarized in Fig.\ \ref{fig:gAgV}.
We obtain preliminary values of 1.10(8) and 1.06(11), for the heavy and light ensembles respectively.
Though the statistics is still low, these values are smaller than the experimental value of 1.2701(25), but are consistent with our earlier lattice calculation \cite{Yamazaki:2008py}, as can be seen in Fig.\ \ref{fig:gAgVmpi2},
where we summarize the dependence on pion mass squared, \(m_\pi^2\).
The calculated values stay away from the experiment as we set the pion mass lighter.
The dependence on the finite-size scaling parameter, \(m_\pi L\), is summarized in Fig.\ \ref{fig:gAgVmpiL}:
\begin{figure}[tb]
\begin{center}
\includegraphics[width=0.75\linewidth,clip]{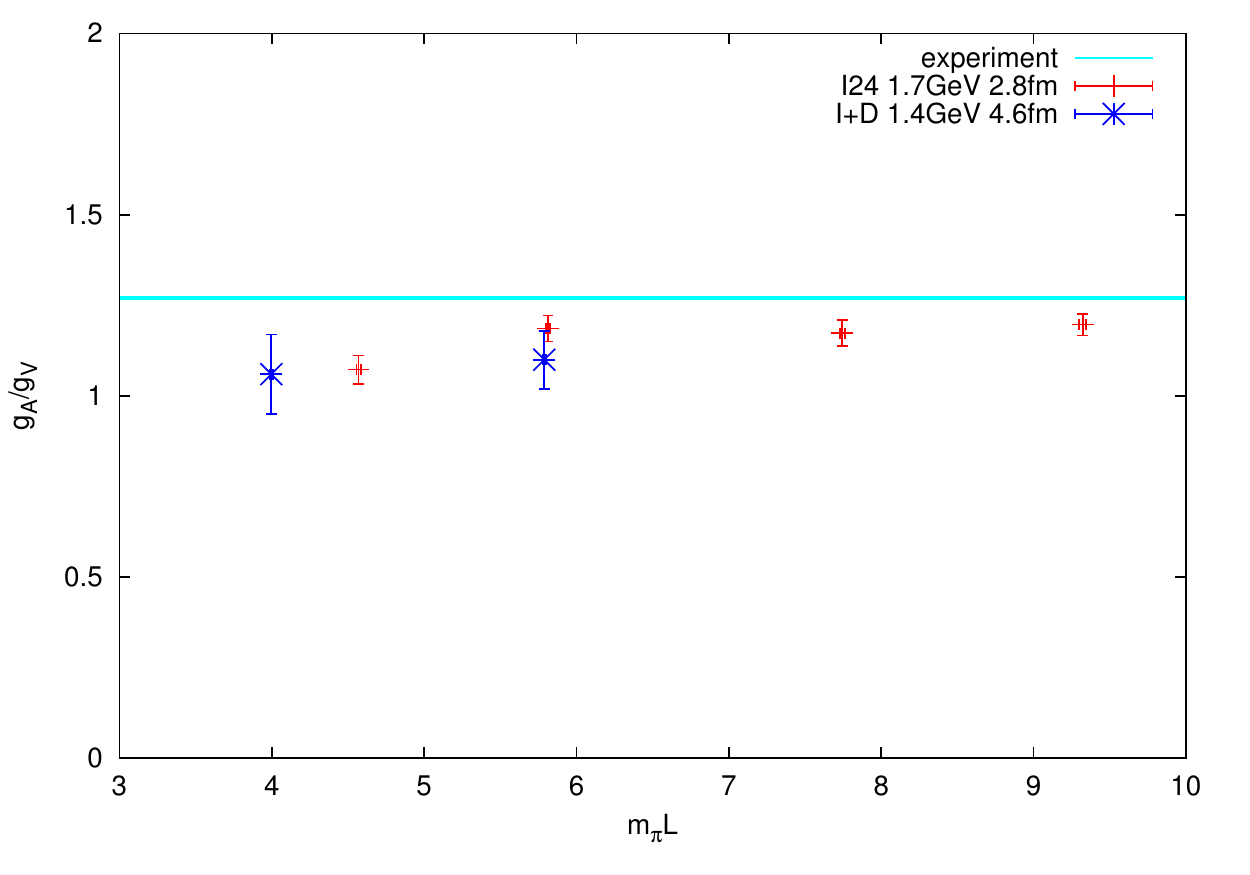}
\end{center}
\caption{
Dependence on a dimensionless finite-size scaling parameter, \(m_\pi L\), of isovector axial to vector charges, \(g_A/g_V\).
The present results are consistent with the scaling discovered in \cite{Yamazaki:2008py}.
}
\label{fig:gAgVmpiL}
\end{figure}
the values from the present 250-MeV and 170-MeV ensembles at \(m_\pi L =\) 5.8 and 4.2, respectively, are consistent with the scaling discovered in \cite{Yamazaki:2008py}.

We also calculate the quark isovector momentum, \(\langle x \rangle_{u-d}\), and helicity, \(\langle x \rangle_{\Delta u - \Delta d}\), fractions and their ratio.
They are still rather noisy at the level of statistics, but seem to confirm that they also share common renormalization.
The ratio, albeit with large statistical error, seem consistent with the experiment.
The individual fractions seem to trend down with decreasing mass as well.

\section{Conclusions}

RBC and UKQCD collaborations continue to work on nucleon structure using the 2+1-flavor dynamical DWF ensembles with lattice cutoff \(\sim\) 1.4 GeV, and \(({\rm 4.6 fm})^3\) spatial volume:
good chiral and flavor symmetries, with residual mass of \(m_{\rm res}a \sim 0.002\), is demonstrated in various observables for the \(m_\pi \sim 170\) and 250 MeV ensembles.
The nucleon mass for these ensembles are estimated at \(m_N \sim 0.98\) and 1.05 GeV, respectively.

Even with the current preliminary low statistics, isovector vector-current form factor are well under control: local-current charge, \(g_V\), agrees well with \(Z_A^{-1}\) from the meson sector.
Axialvector-current form factors are noisier, yet the naturally renormalized ratio, \(g_A/g_V\), yields decent signals and is found consistent with finite-size effect and  scaling in \(m_\pi L\) that were observed earlier \cite{Yamazaki:2008py}.

Moments of structure functions are even noisier, but calculations are well under way:
the naturally renormalized ratio, \(\langle x \rangle_{u-d}/\langle x\rangle_{\Delta u - \Delta d}\), is consistent with experiment.
And the individual fractions are likely showing desired mass dependence.

Our statistics will more than double at least, with the current 8-trajectory interval, and can double further if 4-trajectory interval turns out sufficient in terms of autocorrelation.

%%%%%%%%%%%%%%%%%%%%%%%%%%%%%%%%%%%%%%%%%%%%%%%%
%% BACKMATTER
%%%%%%%%%%%%%%%%%%%%%%%%%%%%%%%%%%%%%%%%%%%%%%%%

I thank RBC and UKQCD Collaborations, especially M.F.~Lin, Y.~Aoki, T.~Blum, C.~Dawson, T.~Izubuchi, C.~Jung, S.~Sasaki and T.~Yamazaki for nucleon calculations, and C.~Kelly and J.~Zanotti for vector and axialvector renormalizations.
I also thank Erlena Dlu for moral support.
RIKEN, BNL, the U.S.\ DOE, University of Edinburgh, and the U.K.\  PPARC provided  facilities essential for the completion of this work.
The I+DSDR ensembles are being generated at ANL Leadership Class Facility (ALCF.)
The nucleon two- and three-point correlators are being calculated at RIKEN Integrated Cluster of Clusters (RICC) and US Teragrid/XSEDE computer clusters.

\end{document}